\documentclass[aps,prl,twocolumn,amsmath,amssymb,floatfix,nofootinbib,10pt]{revtex4}

\usepackage{graphicx,epsfig}
\usepackage{amsmath,amssymb}
\usepackage{subfigure,epsfig}
\usepackage{float}

\newcommand{\etaEW}{\ensuremath{\alpha_{5,\mbox{{\tiny ul}}}}}

\begin{document}

\title{Connection between a possible fifth force and the direct
detection of Dark Matter} \author{Jo Bovy and Glennys R. Farrar}
\affiliation{Center for Cosmology and Particle Physics, Department of
Physics\\ New York University, New York, NY 10003 \\ }

\date{August 12, 2008}

\begin{abstract}
If there is a fifth force in the dark sector and dark sector particles
interact non-gravitationally with ordinary matter, quantum corrections
generically lead to a fifth force in the visible sector. We show how
the strong experimental limits on fifth forces in the visible sector
constrain the direct detection cross section, and the strength of the
fifth force in the dark sector. If the latter is comparable to
gravity, the spin-independent direct detection cross section must
typically be $\lesssim 10^{-55}$ cm$^2$.  The anomalous acceleration
of ordinary matter falling towards dark matter is also constrained:
$\eta_{\rm OM-DM} \lesssim 10^{-8}$.
\end{abstract}

\maketitle

In recent years, the possibility has arisen that there might be a
fifth force acting in the dark sector. (For a theoretical perspective
see, e.g., \cite{Amendola:1999er,Farrar:2003uw} and references
therein.)  A force in the dark sector arises in many of the extensions
of the standard model.  It could be short range (as in most dark
matter models) or long range (through the existence of a very light
boson coupling to dark matter particles); in the latter case it is
known as a ``fifth force".  Light bosons such as the dilaton appear
naturally in supersymmetry and string theory, although if such light
bosons exist they are constrained to couple only very weakly to
ordinary matter thanks to the E\"ot-Wash experiments discussed below.

There are observational motivations to consider a long-range
attractive non-gravitational force between dark matter (DM)
concentrations. Such a force may resolve some discomforts with
conventional $\Lambda$CDM.  If the range of the fifth force is large
compared to the scale of structure formation, it effectively increases
the strength of the gravitational interaction for the dark matter by a
factor $\equiv (1 + \beta)$, and thus would accelerate structure
formation.  This would be helpful because:\\ $\bullet$ Recent
cosmological simulations using a reduced $\sigma_8$ value compatible
with the WMAP year 5 cosmology \cite{2008arXiv0804.2486D}, predict
$z=0$ halo concentrations distinctly lower than the Millennium Run
using the larger WMAP1 value of $\sigma_8$, and find ``very
significant discrepancies with X-ray observations of groups and
clusters of galaxies."  Enhancing the effective gravitational
attraction with a fifth force would mimic the effect of a larger
$\sigma_8$ at recombination and might eliminate this problem. \\
$\bullet$ The number of superclusters observed in SDSS data appears to
be an order of magnitude larger than predicted by $\Lambda$CDM
simulations \cite{2007A&A...462..397E}; accelerated structure
formation would reduce this discrepancy.  \\ $\bullet$ As noted in
\cite{Farrar:2003uw}, a fifth force would tend to clear out the voids;
Ref.~\cite{2005PhRvD..71h3505N} confirms this in a simulation.  This
may improve agreement with $\Lambda$CDM \cite{2001ApJ...557..495P},
although the existence of a discrepancy has been challenged
\cite{2002MNRAS.337.1193M}.\\ $\bullet$ A variety of observations, for
instance the lack of evidence in the Milky Way for a major merger, is
hard to reconcile with the amount of accretion predicted in
$\Lambda$CDM.  Accelerated structure formation reduces late-time
accretion, simply because it leaves less to be accreted later
\cite{2005PhRvD..71h3505N}.  \\ $\bullet$ The number of satellites in
a galaxy such as the Milky Way is predicted to be an order of
magnitude larger than is observed. This ``substructure problem" is
ameliorated by a 5th force, by reducing the stellar content of dwarf
galaxies and making them harder to find.  This is because baryons --
not feeling the 5th force -- are relatively less-bound to dark matter
concentrations than in conventional theory, reducing the amount of
bound gas and lowering the star formation rate in dwarf galaxies, and
increasing the tidal loss of the stars that do form.  However
observational bias may also be responsible for the apparent
contradiction \cite{2008arXiv0806.4381T}.

Experimental constraints on the existence of
equivalence-principle-violating forces for dark matter are relatively
weak \cite{Gradwohl:1992ue}.  On a sub-galactic scale, Kesden and
Kamionkowski pointed out that a difference in acceleration between DM
and stars would change the distribution of stars in the tidal tails of
the Sagittarius dwarf galaxy and claimed a limit about 10\% the
strength of gravity \cite{2006PhRvL..97m1303K}.  However this claim
must be set aside for the time being, because the distances of the
stars in the Stream have been found to have significant systematic
error \cite{2007ApJ...670..346C} and until those distances are
revised, it is difficult to draw any conclusions using the Sagittarius
Stream.  With the original distances there was a difficulty
reconciling the observed line-of-sight velocities of the stars in the
leading stream, which implied a prolate Dark Matter halo, with the
precession of the debris orbital plane, which implied an oblate halo
\cite{2005ApJ...619..807L}.  This discrepancy might be resolved by an
additional attractive force, or may disappear when the stellar
distances are corrected.  On a galaxy-cluster scale, the apparent need
for a fifth force to account for the reported velocity
\cite{2006ESASP.604..723M} of the components of the merging ``bullet"
cluster, IE0657-56, \cite{Farrar:2006tb} has been removed by a better
determination of the velocity \cite{Springel:2007tu}.  Work is
presently underway to extract a limit on or preferred values for the
strength and range of a fifth force from this system (L. Berezhiani
and GRF).

While there is room to consider a fifth force in the dark sector,
constraints on new forces in the visible sector are very
strict. Precise experiments performed by the E\"ot-Wash Group used a
torsion balance to study the differential acceleration of a Be-Ti
test-body pair in the fields of the Earth, the Sun and our Galaxy
\cite{Schlamm}. They tested the universality of free fall, leading
to a limit on a new Yukawa type force between ordinary matter
particles on astronomical length scales
\begin{equation}\label{fifthconstraint}
\alpha_5 \equiv \frac{g_{\mathrm{eff}}^2}{4 \pi u^2 G}\leq 10^{-10} \equiv
\etaEW \ ,
\end{equation}
where $u$ is the atomic mass unit and $g_{\mathrm{eff}}$ is the Yukawa
coupling of ordinary matter to a new light boson. The upper limit
\etaEW\ depends on the specific coupling of the new force, and the
quoted value corresponds to the least stringent limit; more typical
values are two orders of magnitude smaller \cite{Su}. Such a strong
limit arises naturally if some symmetry forbids a tree-level coupling
between ordinary matter and the light boson which mediates the fifth
force, and we assume here that this is the case.  If the tree-level
coupling does not vanish, then our arguments below become even
stronger unless there is extreme fine-tuning.

\begin{figure}[t]
\centering
\subfigure[]{%
\epsfig{figure=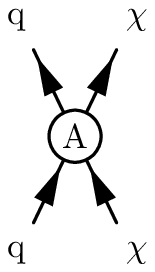}}%
\subfigure[]{%
\epsfig{figure=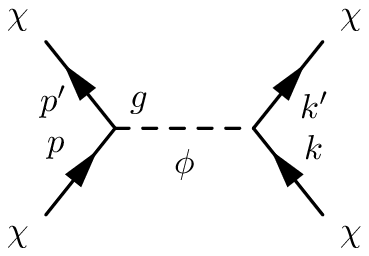}}%
\caption{(a) Direct detection four-fermion vertex coupling quarks and
dark matter fermions $\chi$. (b) Fifth force Yukawa interaction for
fermionic dark matter.}%
\label{directdet}%
\end{figure}

\begin{figure}
\begin{center}
\epsfig{file=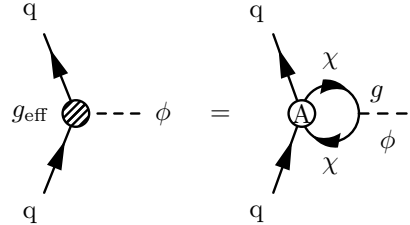}
\caption{An effective Yukawa interaction of ordinary matter particles
arises due to a quantum correction.}\label{quantumcorr}
\end{center}
\end{figure}

In this paper we show that thanks to the bound
(\ref{fifthconstraint}), a connection exists between the strength of
astrophysical effects of a fifth force and laboratory experiments to
detect dark matter.  Namely, the existence of a fifth force between
dark matter particles could set a strong upper bound on the possible
direct detection cross section and vice versa.  From
Figs.~\ref{directdet}(a) and \ref{quantumcorr} it is clear how a fifth
force can arise for ordinary matter as a quantum effect involving a
fifth force between dark matter particles and the direct detection
vertex. 
The constraint (\ref{fifthconstraint}) on the effective Yukawa
interaction between ordinary matter particles can therefore impose a
limit on the product of the DM direct detection cross section and the
DM fifth force, which we determine below.

We assume the fifth force boson to be a light scalar particle,
because a pseudoscalar particle would give a spin-dependent, and thus
non-macroscopic, force, and a vector or axial-vector carrier would give
rise to a repulsive force. If dark matter is fermionic and there is a
non-gravitational attractive fifth force between dark matter
particles, it is most simply modeled as a long-range Yukawa force and
we adopt that description here.  Reality may be more complex.  For
instance in chameleon models \cite{Khoury:2003rn} the new interaction
is damped by sufficiently strong concentrations of matter, so in
principle it could have an impact on structure formation but not be
evident on sub-galactic scales.  The bounds derived here would need to
be revisited in such models.

Consider a Yukawa interaction between a spin-1/2 dark matter field
$\chi$ and a light scalar $\phi$ (spin-0 dark matter is discussed
below)
\begin{equation}\label{lagrangian}
\mathcal{L}_{\mathrm{int}} = - g (\phi-\phi_*) \overline{\chi} \chi \
. 
\end{equation}
Here, $g$ is a dimensionless constant and $g \phi_*$ is the mass of a
dark matter particle in the absence of a vacuum expectation value
$\langle \phi \rangle$ of the scalar field.

%EFFECTIVE YUKAWA INTERACTION FIGURE

The potential corresponding to this interaction is found by comparing
the amplitude of the process in Fig.~\ref{directdet}(b) to the Born
approximation formula from non-relativistic quantum mechanics: the
scattering amplitude in the low-energy approximation, normalized
according to non-relativistic conventions is given by
\begin{equation}\label{ampli}
\mathcal{M} = \frac{g^2}{|\mathbf{q}|^2+m_\phi^2} \qquad (\mathbf{q} =
\mathbf{p}'-\mathbf{p}) \ ,
\end{equation}
which is simply the Fourier transform of the potential, giving the
familiar Yukawa potential (see, e.g., \cite{Peskin:1995ev})
\begin{equation}\label{potential}
V(r) = -\frac{g^2}{4\pi}\frac{1}{r} e^{-m_\phi r} \ ,
\end{equation}
in which $m_\phi$ is the mass of the scalar.  At distances small
compared to the range of the force, $m_\phi^{-1}$ in natural units,
the force is just proportional to the gravitational attraction, with
proportionality constant $\beta$:
\begin{equation}\label{ratio}
\beta \equiv \frac{g^2/4 \pi}{G m_{\chi}^2} = \frac{1}{4 \pi G
(\langle \phi \rangle -\phi_*)^2} = \frac{m_{\mathrm{Pl}}^2}{4 \pi (
\langle \phi \rangle - \phi_*)^2}\ ,
\end{equation}
since the mass of the dark matter particle is $m_{\chi} = g(\langle
\phi \rangle-\phi_*)$.  Notice how $g$ drops out of equation
(\ref{ratio}), leaving the relative strength of the fifth force
compared to gravity and the mass of the dark matter particles as two
independent quantities. This relationship, derived here from the
single particle perspective, was found earlier from density
perturbation considerations \cite{Farrar:2003uw}.

A few observations: a) In order for the fifth force to be of the order
of the gravitational force, i.e. to be astrophysically relevant,
$\langle \phi \rangle - \phi_*$ should be of the order of the Planck
mass. This choice is quite natural, since there is no immediate
relation between the scalar and any of the lower mass scales. Since
$\langle \phi \rangle - \phi_*$ appears in the denominator, a smaller
value would mean a much larger fifth force. b) With $\langle \phi
\rangle - \phi_*$ of the order of the Planck scale, the dark matter
would be very heavy in most scenarios. To get a mass in the
electroweak region would require $g, \tilde{g}$ $\sim$
$O(m_{\mathrm{W}}/m_{\mathrm{Pl}})$, which presumably should be
enforced by some symmetry. For $g, \tilde{g}$ closer to unity, we can
infer that dark matter must be a nonthermal relic. Nonthermal relics
are produced in sufficient abundance if their mass is of order the
inflaton mass (presumably around $10^{13}$ GeV)
\cite{Chung:1998zb}. This would connect fifth force physics with
inflation physics, which would be interesting, but not within the
scope of this work.

In direct detection experiments, evidence for dark matter is sought in
the observation of ordinary matter particles recoiling due to their
being scattered by dark matter particles. The most general
non-derivative four-fermion operator is given by
\begin{equation}\label{matrixelement}
\mathcal{M} = \sum_{i} A_{i} ( \overline{q} O_{i}
q)(\overline{\chi}O_i \chi) \ ,
\end{equation}
where the sum is over scalar (S), pseudoscalar (P), vector (V),
axial-vector (A), and tensor vertices (T), and the $A_i$ are constants
with dimensions of energy$^{-2}$. If the DM particle is Majorana, the
vector and tensor terms are absent in (\ref{matrixelement}); this is
commonplace in supersymmetric theories.
In the limit of small momenta, the only contributing terms are the
scalar, vector, axial-vector, and tensor interactions. These last two 
correspond to the spin-dependent cross section, which the reasoning
below will fail to limit.

The spin-independent (SI) cross section for (fermionic) DM - quark
scattering is
\begin{equation}\label{dd}
\sigma^{\mathrm{SI}}_{q \chi}
 = (A_S + A_V)^2 \frac{m_q^2}{\pi} \ ,
\end{equation}
in the limit that the DM mass is much larger than $m_q$ (the effective
quark mass, $\sim \Lambda_{\mathrm{QCD}}$).

Figure \ref{quantumcorr} shows how an effective Yukawa-type fifth
force interaction of quarks arises as a consequence of a quantum
correction.  The correction is made up of two parts: first, the direct
detection vertex couples the two quarks to two virtual dark matter
particles; second, these dark matter particles couple to the fifth
force scalar.  For scalar $\phi$ the scalar and tensor contributions to the sum
(\ref{matrixelement}) give non-vanishing effective couplings
\begin{equation}\label{geff1}
g_{\mathrm{eff}} \approx A_{S,T} \,g \, \Lambda^2  \ ,
\end{equation}
where $\Lambda$ is the ultraviolet cut-off, since this diagram would
be quadratically divergent if the 4-fermion vertex were pointlike. In
reality, we expect the short distance theory to be renormalizable,
giving the effective 4-fermi vertex a form-factor.  This cuts off the
divergence at some model-dependent scale leading to an expression like
(\ref{dd}) with $\Lambda$ for instance of the order of the TeV-scale,
the dark matter-scale, or the Planck-scale.

Through an insertion of the Higgs vacuum expectation value into the
loop diagram, the V
contribution to the sum (\ref{matrixelement}) gives
a non-vanishing effective vertex
\begin{equation}\label{geffV}
g_{\mathrm{eff}} \approx A_V g\,  m_q m_\chi \ln \Lambda 
\ ,
\end{equation}
since this diagram is now superficially logarithmically divergent. The
 quantum correction in Fig.~\ref{quantumcorr} from the axial-vector
 term in (\ref{matrixelement}) vanishes. Therefore, $A_A$ is not
 constrained and no relation between a fifth force for DM and a fifth
 force for ordinary matter ensues.

Through (\ref{ratio}) we re-express the dark matter Yukawa-coupling as
$g^2 = 4 \pi \beta G m_{\chi}^2$,
where $\beta$, the relative strength of the fifth force with respect
to gravity, is $\lesssim O(1)$ from astrophysics
\cite{Gradwohl:1992ue}. Imposing the experimental limit
(\ref{fifthconstraint}) on the effective vertex (\ref{geff1}), leads
to
\begin{equation}\label{Aconst}
A_{S,T}^2 \lesssim \frac{\etaEW}{\beta} \frac{u^2}{m_{\chi}^2 \Lambda^4} \ ,
\end{equation}
or for the vector coupling
\begin{equation}\label{AconstV}
A_V^2 \lesssim \frac{\etaEW}{\beta} \frac{u^2}{m_{\chi}^4 m_q^2 \ln^2 \Lambda}
\ .
\end{equation}

Since the bounds on $A_i$ become ever more stringent for higher DM
mass scales, we use the smallest $m_{\chi}$, i.e., weak scale DM, to
obtain the most conservative limits.  If both $A_S$ or $A_T$ and $A_V$ are
present in (\ref{dd}), the constraint (\ref{AconstV}) is dominant and
from (\ref{dd}) and (\ref{AconstV}) we find an upper bound on the
product of $\beta$ and the spin-independent direct detection cross
section:
\begin{align}
\beta \Bigg(\frac{\sigma^{\mathrm{SI}}_{q \chi}}{10^{-43} \
\mathrm{cm}^2} \Bigg) \lesssim \, 10^{-3} \times &
\Bigg(\frac{\etaEW}{10^{-10}}\Bigg) \label{crossbound2}\\ &
\Bigg(\frac{100 \ \mathrm{GeV}}{m_\chi}\Bigg)^4 \Bigg(\frac{1}{\ln
\Lambda}\Bigg)^2 .\notag
\end{align}
However, if only the scalar term in (\ref{matrixelement}) is present
in (\ref{dd}) -- e.g., because DM is Majorana -- then (\ref{Aconst})
leads to the more stringent bound:
\begin{align}
\beta \Bigg(\frac{\sigma^{\mathrm{SI}}_{q \chi}}{10^{-43}
\ \mathrm{cm}^2}\Bigg)  \lesssim  \,10^{-12} \times
& \Bigg(\frac{\etaEW}{10^{-10}}\Bigg) 
\Bigg(\frac{m_q}{\Lambda_{\mathrm{QCD}}}\Bigg)^2 \label{crossbound}\\ & \Bigg(\frac{100 \ \mathrm{GeV}}{m_\chi}\Bigg)^2 \Bigg(\frac{1 \
\mathrm{TeV}}{\Lambda}\Bigg)^4\ . \notag
\end{align}

The product of the spin-dependent (SD) cross section and $\beta$ is
similarly constrained in the case that only a tensor term contributes
to the matrix element (\ref{matrixelement}). The SD cross section is
like in (\ref{dd}) but with $(A_S + A_V)^2$ replaced by $6A_T^2$. The
bound on the product of the SD cross section and $\beta$ is then of
the same form as equation (\ref{crossbound}), but with the coefficient
$10^{-11}$.

In the case of spin-0 dark matter, we take the interaction Lagrangian
to be 
$\mathcal{L}_{\mathrm{int}} =  - \tilde{g}  m_{\chi} (\phi-\phi_*) \phi_\chi^2$,
in which $\phi_\chi$ is the spin-0 dark matter field and the coupling
$\tilde{g}$ is chosen so that the amplitude of the process in
Fig.~\ref{directdet}(b) is again given by (\ref{ampli}), and
consequently the potential is the same as for fermionic DM, with $g
\rightarrow \tilde{g}$.  The spin-0 DM mass is $m_{\chi} =
2\tilde{g}(\langle \phi \rangle-\phi_*)$,
so the relative strength of the fifth force with respect to gravity at
distances $\ll m_\phi^{-1}$ in the spin-0 case is then
$\tilde{\beta} = \frac{m_{\mathrm{Pl}}^2}{16 \pi ( \langle \phi \rangle - \phi_*)^2}$ .
We parametrize the direct detection vertex as
$\mathcal{M} = \tilde{A_S} m_{\chi}\overline{q} q \phi_{\chi}^2$ ,
leading to the DM-quark cross section
$\sigma^{\mathrm{SI}}_{q \chi} = \tilde{A_S}^2 m_q^2/4 \pi$.
The diagram in Figure \ref{quantumcorr} now 
is superficially logarithmically divergent, which leads us to write
$g_{\mathrm{eff}} \approx \tilde{A_S} \tilde{g} m_{\chi}^2 \ln \Lambda
$.
As before we get an upper bound on the four-fermion vertex: 
$\tilde{A_S}^2 \lesssim \frac{\etaEW}{\tilde{\beta}}
\frac{u^2}{m_{\chi}^6 \ln^2 \Lambda}$, 
which leads to
\begin{align}
\beta \Bigg(\frac{\sigma^{\mathrm{SI}}_{q \chi}}{10^{-43} \
\mathrm{cm}^2} \Bigg) \lesssim \, 10^{-9} & \times
\Bigg(\frac{\etaEW}{10^{-10}}\Bigg)
\Bigg(\frac{m_q}{\Lambda_{\mathrm{QCD}}}\Bigg)^2 \label{crossbound3}
\\ & \times
\Bigg(\frac{100 \ \mathrm{GeV}}{m_\chi}\Bigg)^6 \Bigg(\frac{1}{\ln
\Lambda}\Bigg)^2 \ ,\notag
\end{align}
for scalar DM.

If there is an astrophysically relevant $\beta$, these bounds on the
spin-independent direct detection cross section are smaller than
current experimental limits, which restrict the SI scattering cross
section of DM on ordinary matter to be less than
$10^{-43}$ cm$^2$ for $m_\chi = $ 100 GeV and less than $10^{-30}$
cm$^2$ for superheavy DM, $m_\chi = $ $10^{13}$ GeV
\cite{Angle:2007uj}.

Before closing, we note that the effective vertex in Fig.~2 discussed
above also leads to a fifth force between ordinary matter (OM) and
dark matter proportional to 
$g g_{\mbox{eff}}$.  This implies that the fractional anomalous
acceleration of two materials falling towards DM is given
by
\begin{equation}
\eta_{\rm OM-DM} \lesssim 10^{-8} \times \Bigg(\frac{\eta_{\rm OM-OM}
\times \Delta(q/\mu)\times \beta}{\ \ 10^{-13} \ \ \times \ 10^{-3} \
\ \times 1}\Bigg)^{1/2} \ ,
\end{equation}
where $\eta_{\rm OM-OM}$ is the fractional anomalous acceleration for
the two materials falling towards OM and $\Delta(q/\mu)$ is the
difference of the ratio of the fifth-force-charge to the mass of the
different materials, with the average $q/\mu$ of the sun equal to one.
This limit is more stringent than current experimental bounds,
$\eta_{\rm OM-DM} \lesssim 10^{-5}$ \cite{Schlamm}.

To summarize, we have shown here that if there is a cosmologically
important dark matter fifth force, then the experimental limit on a
fifth force for ordinary matter essentially excludes the direct
detection of dark matter particles in the near future, except through
the spin-dependent cross section. Generically, for weak scale dark
matter and a gravitational-strength fifth force, 
$\sigma^{\mathrm{SI}}_{q \chi} \leq 10^{-55}$ cm$^2$; in the least
constrained case the cross section must be $\leq 10^{-46}$ cm$^2$. For
heavier dark matter, as in nonthermal relics scenarios, the bound is
even smaller. Conversely, if the spin-independent interaction of dark
matter is observed in laboratory experiments, our argument shows that
any fifth force in the dark sector must be too weak to be
astrophysically relevant.  The tensor part of the spin-dependent cross
section is similarly constrained.  We have also given a model
independent upper bound on a non-gravitational interaction between
ordinary matter and dark matter, which is significantly better than
current direct experimental limits.

{\bf Acknowledgments:} The research of GRF has been supported in part
by NSF grants PHY-0245068, PHY-0701451 and NASA NNX08AG70G.

\end{document}